\documentclass[preprint,12pt]{elsarticle}

\usepackage{amssymb,amsfonts}

\usepackage{multirow}

\usepackage{hyperref}
\hypersetup{hidelinks}

\usepackage{cellspace}
\usepackage{hhline}

\usepackage{stackengine}
\setstackEOL{\\}

\usepackage{lineno}

\usepackage{multirow}

\usepackage{array}
\newcolumntype{M}[1]{>{\centering\arraybackslash}m{#1}}

\usepackage{tabularx}

\usepackage{lscape}

\usepackage[table]{xcolor}

\usepackage{longtable}

\usepackage{booktabs, makecell}

\usepackage{pifont}
\newcommand{\cmark}{\ding{51}}
\newcommand{\xmark}{\ding{55}}
\newcommand{\qmark}{\texttt{?}}
\newcommand{\comment}[1]{}
\newcommand{\NA}{N/A}

\journal{arXiv}

\begin{document}

\begin{frontmatter}

\title{Applying data technologies to combat AMR: current status, challenges, and opportunities on the way forward}

\author[inst1]{Leonid Chindelevitch\fnref{jointfirst}\corref{corr1}}

\ead{l.chindelevitch@imperial.ac.uk}

\author[inst1,inst2]{Elita Jauneikaite\fnref{jointfirst}\corref{corr1}}

\ead{e.jauneikaite@imperial.ac.uk}

\fntext[joinedfirst]{These authors contributed equally to this work.}
\cortext[corr1]{Corresponding authors}

\author[inst5]{Nicole E. Wheeler}

\author[inst14]{Kasim Allel}

\author[inst25]{Bede Yaw Ansiri-Asafoakaa}

\author[inst6]{Wireko A. Awuah}

\author[inst9,inst11]{Denis C. Bauer}

\author[inst10]{Stephan Beisken}

\author[inst12]{Kara Fan}

\author[inst18]{Gary Grant}

\author[inst16,inst3]{Michael Graz}

\author[inst21]{Yara Khalaf}

\author[inst19]{Veranja Liyanapathirana}

\author[inst13]{Carlos Montefusco-Pereira}

\author[inst7]{Lawrence Mugisha}

\author[inst8]{Atharv Naik}

\author[inst20]{Sylvia Nanono}

\author[inst85]{Anthony Nguyen}

\author[inst2,inst27]{Timothy Rawson}

\author[inst22]{Kessendri Reddy}

\author[inst0]{Juliana M. Ruzante}

\author[inst24]{Anneke Schmider}

\author[inst17]{Roman Stocker}

\author[inst1]{Leonhardt Unruh}

\author[inst15]{Daniel Waruingi}

\author[inst3]{Heather Graz}

\author[inst4]{Maarten van Dongen}

\address[inst1]{MRC Centre for Global Infectious Disease Analysis, Imperial College London,
            Praed Street, 
            London,
            W2 1NY, 
            England,
            UK}

\address[inst2]{NIHR HPRU in Healthcare Associated Infections and Antimicrobial Resistance, Imperial College London
            Du Cane Road, 
            London,
            W12 0NN, 
            England,
            UK}
            
\address[inst5]{Institute of Microbiology and Infection, University of Birmingham,
            Edgbaston, 
            Birmingham,
            B15 2TT, 
            England,
            UK}
            
\address[inst14]{Department of Disease Control, London School of Hygiene and Tropical Medicine,
            Keppel Street, 
            London,
            WC1E7HT, 
            England,
            UK}

\address[inst25]{Royal Caleb Pharmacy,
            22 Pampaso Road, 
            Kumasi,
            Ghana}
            
\address[inst6]{Sumy State University,
            2 Rymskogo-Korsakova St.,
            Sumy,
            40007,
            Ukraine}

\address[inst9]{CSIRO,
            Building 53, 11 Julius Ave, 
            North Ryde,
            2113,
            New South Wales,
            Australia}

\address[inst11]{Departments of Biomedical Sciences and Applied BioSciences, Macquarie University,
            205B Culloden Rd, 
            Sydney,
            2109, 
            New South Wales,
            Australia}            

\address[inst10]{Ares Genetics GmbH,
            Karl-Farkas-Gasse 18, 
            Vienna,
            1030, 
            Austria}
            
\address[inst12]{Westview High School,
            13500 Camino Del Sur,
            San Diego,
            92129, 
            CA,
            USA}
            
\address[inst18]{School of Pharmacy and Medical Sciences, Gold Coast campus, Griffith University,
            Parklands Drive, 
            Southport,
            4222, 
            Queensland,
            Australia}
            
\address[inst16]{School of Law, Univeristy of Bristol,
            Wills Memorial Building, Queens Road, 
            Clifton, Bristol,
            England,
            BS8 1RJ, 
            UK}
            
\address[inst3]{Biophys Ltd,
            4 Four Ash Court, Usk, 
            Monmouthshire,
            NP15 1BE, 
            Wales,
            UK}
            
\address[inst21]{Department of Epidemiology, High Institute of Public Health, Alexandria University,
            165 El-Horreya Road, 
            Bab Sharqi,
            6W4G+7R, 
            Alexandria,
            Egypt}
            
\address[inst19]{Department of Microbiology, Faculty of Medicine, University of Peradeniya,
            Galaha Road, 
            Peradeniya, Kandy,
            20400,
            Sri Lanka}
            
\address[inst13]{Boehringer Ingelheim Pharma, Pharmaceutical Development Biologicals,
            65, Birkendorfer Strasse, 
            Biberach an der Riss,
            88400, 
            Baden-Württemberg,
            Germany}
            
\address[inst7]{Department of Wildlife and Aquatic Animal Resources (WAAR), College of Veterinary Medicine, Animal Resources and Biosecurity (COVAB), Makerere University,
            P.O.Box 7062, 
            Kampala,
            Uganda}

\address[inst8]{Department of Bioengineering, Imperial College London,
            Exhibition Road, 
            London,
            SW7 2AZ, 
            England,
            UK}
            
\address[inst20]{Infectious Diseases Institute, College of Health Sciences, Makerere University,
            University Road, 
            Kampala,
            22418, 
            Uganda}
            
\address[inst85]{The Australian e-Health Research Centre, CSIRO,
            Level 7, Surgical Treatment and Rehabilitation Service, 296 Herston Road, 
            Herston,
            QLD 4029,
            Queensland,
            Australia}
            
\address[inst27]{Centre for Antimicrobial Optimisation, Hammersmith Hospital, Imperial College London,
            Du Cane Road, 
            London,
            W12 0NN,
            England,
            UK}
            
\address[inst22]{Division of Medical Microbiology, Faculty of Medicine and Health Sciences, Stellenbosch University/National Health Laboratory Service Tygerberg,
            Francie Van Zyl Avenue, 
            Parow,
            7500, 
            Cape Town, South Africa}

\address[inst0]{RTI International,
            3040 East Cornwallis Road, 
            Research Triangle Park,
            27709-2194, 
            NC,
            USA}
            
\address[inst24]{Chatham House Centre for Global Health,
            10 St James's Square, 
            SW1Y 4LE,
            London,
            England,
            UK}
            
\address[inst17]{PhAST Diagnostics,
            38 Wareham St., Fl. 2,
            Boston,
            02118, 
            MA,
            USA}

\address[inst15]{Students Against Superbugs Africa,
             Nairobi,
            Kenya}
            
\address[inst4]{AMR Insights,
            Keizersgracht 482, 
            Amsterdam,
            1017 EG, 
            Noord-Holland,
            Netherlands}

\begin{abstract}
Antimicrobial resistance (AMR) is a growing public health threat, estimated to cause over 10 million deaths per year and cost the global economy 100 trillion USD by 2050 under status quo projections. These losses would mainly result from an increase in the morbidity and mortality from treatment failure, AMR infections during medical procedures, and a loss of quality of life attributed to AMR. Numerous interventions have been proposed to control the development of AMR and mitigate the risks posed by its spread.\hfill\break\break
This paper reviews key  aspects of bacterial AMR management and control which make essential use of data technologies such as artificial intelligence, machine learning, and mathematical and statistical modelling, fields that have seen rapid developments in this century. Although data technologies have become an integral part of biomedical research, their impact on AMR management has remained modest. We outline the use of data technologies to combat AMR, detailing recent advancements in four complementary categories: surveillance, prevention, diagnosis, and treatment. We provide an overview on current AMR control approaches using data technologies within biomedical research, clinical practice, and in the ``One Health'' context. We discuss the potential impact and challenges wider implementation of data technologies is facing in high-income as well as in low- and middle-income countries, and recommend concrete actions needed to allow these technologies to be more readily integrated within the healthcare and public health sectors.\hfill\break
\end{abstract}


\end{frontmatter}

\section*{The future of AMR control}

\noindent Let us start by imagining what the future of treating and preventing infections caused by AMR could look like.\hfill\break\break
\noindent You get up in the morning; your watch uses its smart sensors to measure your temperature, pulse and heart rate, and monitors your health through its built-in microfluidic device. When a possible infection is suspected due to an increased body temperature and the fact that you got up several times during the night, an alert comes up on your phone prompting you to provide a urine sample that you put in a previously distributed electronically labelled tube, which gets collected by a dedicated drone.\hfill\break\break
By the time you arrive to work, your phone has sent you a confirmation that you have a bacterial infection, so you are allowed to come in, but that it is a Gram-negative bacterium from a urinary tract infection and you need to start a course of antibiotic treatment as soon as possible once it is identified.\hfill\break\break
Before lunch break you get another alert informing you that the bacterium is \textit{E. coli} and fortunately, it is a strain susceptible to trimethoprim, which your family doctor prescribes you. At lunchtime your pharmacist has the tablets discreetly delivered to you via drone, with an electronic certificate ensuring they are genuine. You start a course of treatment immediately after lunch, and your family doctor gets informed via a confidential, encrypted application on your phone. You complete the treatment; a follow-up test confirms you cleared the infection and your health is back to normal.\hfill\break\break
Your information gets automatically and anonymously added to a database tracking the geographically located prevalence and incidence of AMR; thanks to the whole-genome sequencing of your strain, public health authorities link it to an ongoing outbreak and track its transmission routes, which leads to a community-based information campaign that helps to stop the outbreak. \hfill

\tableofcontents

\section{Introduction}

\subsection{Overview of the challenges of AMR}

\begin{figure}[ht]
    \includegraphics[width=0.95\textwidth]{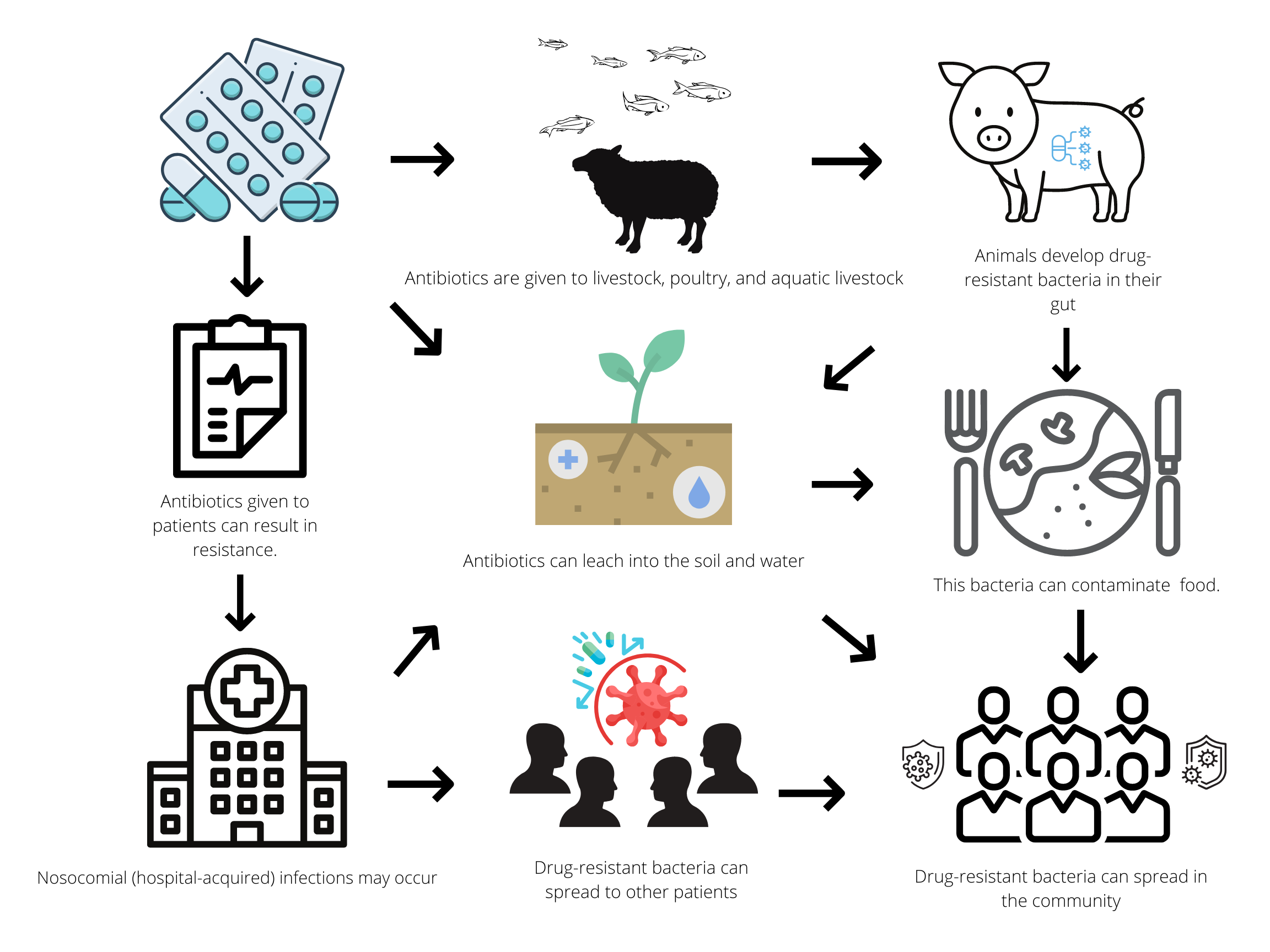}
    \caption{Overview of interactions between different spheres facilitating antimicrobial resistance dissemination. Figure adapted from the WHO Global Action Plan on AMR \cite{TriAgency}.}
    \label{F:Spheres}
\end{figure}

\noindent Antimicrobial resistance (AMR), an ability of bacteria to escape the action of antimicrobial treatment, has become a major public health problem, projected to cause over 10 million human deaths annually by 2050 \cite{ONeill}. A comprehensive systematic study estimated that globally AMR was associated with 4.95 million deaths, including the direct contribution to 1.27 million deaths, in 2019 \cite{AMRAssessment}. This problem will likely be exacerbated by the COVID-19 pandemic that swept the world and resulted in sudden and inconsistent changes in antimicrobial prescription and stewardship practices globally \cite{hsu2020covid}. \hfill\break\break
The World Health Organisation (WHO) and the US Center for Disease Control (CDC) have both compiled a list of clinically important bacterial pathogens that require urgent actions to address their high levels of antibiotic resistance and global spread \cite{WHOPriority, CDCList}. In addition, a priority list was compiled by the WHO for India last year \cite{IPPL}. The three pathogen categories specified by the WHO are critical, high and medium priority groups, with a separate mention for multidrug resistant (MDR) and extensively drug resistant (XDR) \textit{M. tuberculosis}; the CDC subdivides the pathogens into urgent, serious, and concerning threats, and places additional ones on a Watch List. \hfill\break\break
Many of the pathogens on the combined list (Table \ref{T:WHO}) are able to infect both humans and animals. Reducing the prevalence of those pathogens in animals, food, and the environment can reduce human exposure and reduce infections that will need to be treated. Antimicrobials are used in animals and food production to treat and prevent disease, and sometimes as growth promoters. However, this use of antimicrobials contributes to the development of AMR in both pathogenic and commensal bacteria that can spread from the environment via food, water and soil (Figure \ref{F:Spheres}). This highlights the need for unified AMR policies set out in One Health terms.\hfill

\setlength\extrarowheight{3pt}
\begin{longtable}[c]{|M{0.32\linewidth}|M{0.23\linewidth}|M{0.14\linewidth}|M{0.15\linewidth}|M{0.11\linewidth}|}
    \hline \textbf{Bacterial species of concern} & \textbf{Characteristic} & \textbf{WHO Priority level} & \textbf{CDC Threat level} & \textbf{India Priority level} \\[5pt]
\hline
\endfirsthead
\textit{Mycobacterium tuberculosis}             & Drug-resistant        & Separate category         & Serious &  \NA \\
\hline
\multirowcell{2}{\textbf{\textit{Acinetobacter}}\\ \textbf{\textit{baumannii}}}                & Carbapenem-resistant  & Critical  & Urgent        & \multirowcell{3}{Critical} \\ \hhline{|~|-|-|-|~}
                                                & Colistin-resistant    &  \NA                            & \NA             & \\
\hline
\multirowcell{4}{\textbf{\textit{Pseudomonas}}\\\textbf{\textit{aeruginosa}}}  & Carbapenem-resistant  & Critical                  & \multirowcell{2}{\NA}       & \multirowcell{2}{Critical} \\\hhline{|~|-|-|~|~}
                                                & Colistin-resistant    & \multirowcell{2}{\NA}                             &              &\\\hhline{|~|-|~|-|-}
                                                & Multidrug-resistant   &                        & Serious   & \NA\\ 
\hline
\multirowcell{4}{\textbf{\textit{Klebsiella}} \\\textbf{\textit{pneumoniae}}\\\textbf{\textit{Escherichia coli}}}    &                                                         Carbapenem-resistant & \multirow{2}*{Critical}   & Urgent   & Critical\\
\hhline{|~|-|~|-|-|}
                                                & ESBL-producing        &                           & Serious   & \NA\\
\hhline{|~|-|-|-|-|}
                                                & Tigecycline-resistant & \multirowcell{2}{\NA}                          & \multirowcell{2}{\NA}          & \multirowcell{2}{Critical}\\
\hhline{|~|-|~|~|~}
                                                & Colistin-resistant    &                              &              &\\
\hline
\multirow{2}*{\textbf{\textit{Enterobacter spp.}}}       & Carbapenem-resistant  & \multirow{2}*{Critical}   & Urgent    & \multirow{2}*{\NA} \\
\hhline{|~|-|~|-|~|}
                                                & ESBL-producing        &                           & \NA       & \\ 
\hline
\multirow{2}*{\setstackgap{L}{2ex}\Centerstack{\textbf{\textit{Serratia spp.}}\\\textbf{\textit{Proteus spp.}}\\\textbf{\textit{Providencia spp.}}\\\textbf{\textit{Morganella spp.}}}}    
                                                & Carbapenem-resistant  & \multirow{2}*{Critical}   & \multirow{2}*{\NA}    & \multirow{2}*{\NA} \\
\hhline{|~|-|~|~|~|}
                                                & ESBL-producing        &                           &           & \\ 
\hline
\multirow{3}*{\textbf{\textit{Enterococcus spp.}}}                     & Vancomycin-resistant & High (\textit{E. faecium}) & Serious (\textit{E. faecium)} & \multirowcell{3}{High} \\
\hhline{|~|-|-|-|~|}
                                                                       & Linezolid-resistant    &    \multirowcell{2}{\NA}                    & \multirowcell{2}{\NA}                      &   \\\hhline{|~|-|~|~|~|}
                                                                       & Daptomycin non-susceptible &                   &                       &   \\
\hline
\multirowcell{4}{\textbf{\textit{Staphylococcus}}\\\textbf{\textit{aureus}}}   & Methicillin-resistant & \multirow{2}*{High}       & Serious   & \multirowcell{4}{High} \\
\hhline{|~|-|~|-|~|}
                                                & Vancomycin-intermediate and resistant &           & \multirowcell{3}{\NA}        & \\\hhline{|~|-|-|~|~|}
                                                & Linezolid-resistant                            & \multirowcell{2}{\NA}  &  & \\\hhline{|~|-|~|~|~|}
                                                & Daptomycin non-susceptible            &       &        &  \\
                                                
\hline
\multirowcell{2}{\textbf{\textit{Staphylococcus},}\\\textbf{\textit{coagulase-negative}}} & Vancomycin-resistant & \multirowcell{2}{\NA} & \multirowcell{2}{\NA} & \multirowcell{2}{Medium}\\\hhline{|~|-|~|~|~|}
& Linezolid-resistant & & & \\
\hline
\multirow{2}*{\textbf{\textit{Campylobacter}}}           & Fluoroquinolone-resistant & High                  & \multirow{2}*{Serious}   & \multirow{2}*{\NA} \\ 
\cline{2-3} 
                                                & Macrolide-resistant       & \NA                   &    & \\ 
\hline   
\multirow{4}*{\textbf{\textit{Salmonella spp.}}}               & Fluoroquinolone-resistant & High                  & \multirowcell{4}{Concerning}          & \NA \\ 
\hhline{|~|-|-|~|-|}
                                                & 3rd generation Cephalosporin-resistant & \multirowcell{3}{\NA}           & & \multirowcell{3}{High} \\
\hhline{|~|-|~|~|~}
                                                & Azithromycin-resistant     &                                 &                       &   \\
\hhline{|~|-|~|~|~}
                                                & Carbapenem-resistant       &  &   &   \\
\hline
\textbf{\textit{Helicobacter pylori}}           & Clarithromycin-resistant  & High                  &       \NA  & \NA \\ 
\hline
\multirow{3}*{\textit{Neisseria gonorrhoeae}}   & 3rd generation Cephalosporin-resistant & \multirow{2}*{High} & \multirow{3}*{Urgent} & \multirow{3}*{\NA} \\ 
\hhline{|~|-|~|~|~|}
                                                & Fluoroquinolone-resistant                 &                     & &                       \\
\hhline{|~|-|-|~|~|}                                                
                                                & Resistant to other drugs                  &  \NA                   &  & \\ 
\hline
\multirow{2}*{\textit{Neisseria meningitidis}}           & Fluoroquinolone-non-susceptible          & \multirowcell{2}{\NA} & \multirowcell{2}{\NA} & \multirowcell{2}{Medium} \\\hhline{|~|-|~|~|~|}
                                                & Third generation cephalosporin-non-susceptible & & &\\
\hline
\multirowcell{5}{\textit{Streptococcus} \\ \textit{pneumoniae}} & Penicillin-non-susceptible & Medium & \multirowcell{5}{Serious} & \NA\\
\hhline{|~|-|-|~|-|}
                                                                & Cephalosporin-resistant   & \multirowcell{4}{\NA} & & \multirowcell{3}{Medium}\\
\hhline{|~|-|~|~|~|}
                                                                & Fluoroquinolone-resistant   & & &\\
\hhline{|~|-|~|~|~|}
                                                                & Linezolid-resistant   & & &\\
\hhline{|~|-|~|~|-|}                                                               
                                                                & Resistant to other drugs                  &                  &  & \NA \\ 
\hline                                                                
\hline
\multirow{3}*{\textit{Haemophilus influenzae}}             & Ampicillin-resistant      & Medium & \multirowcell{3}{\NA}      & \NA \\\hhline{|~|-|-|~|-|}
                                            & Third generation cephalosporin-non-susceptible & \multirowcell{2}{\NA} & & \multirowcell{2}{Medium} \\\hhline{|~|-|~|~|~|}
                                            & Carbapenem-non-susceptible & & & \\
\hline
\multirow{4}*{\textit{Shigella spp.}}       & Fluoroquinolone-resistant & Medium & \multirow{2}*{Serious} & \multirow{2}*{\NA}\\\hhline{|~|-|-|~|~|}
                                            & Macrolide-resistant       & \multirowcell{3}{\NA}    &  & \\\hhline{|~|-|~|-|-|}
                                            & Third generation cephalosporin-resistant & & \multirowcell{2}{\NA} & \multirowcell{2}{Medium} \\\hhline{|~|-|~|~|~}
                                            & Azithromycin & & & \\
\hline
\textbf{\textit{Clostridioides difficile}}           & Drug-resistant            & \NA    & Urgent & \NA \\
\hline
\textit{Group A Streptococcus}        & Erythromycin-resistant    &  \NA      &  Concerning  & \NA \\
\hline
\textbf{\textit{Group B Streptococcus}}        & Clindamycin-resistant     & \NA       &  Concerning  & \NA \\
\hline
\textit{Mycoplasma genitalium}              & Drug-resistant            & \NA       &   Watch List  & \NA \\
\hline
\textit{Bordetella pertussis}               & Drug-resistant            & \NA       &   Watch List  & \NA   \\
\hline
\caption{Annotated list of bacterial pathogens according to WHO and Indian priority and CDC threat levels; a line between two drug classes means the bacteria must be resistant to both for the WHO classification, but either one for the other classifications. Pathogens in bold can also infect animals.}
\label{T:WHO}
\end{longtable}

\noindent In recent decades, a number of initiatives have been established globally to support national and international actions to be taken against AMR by individual countries and beyond \cite{quadripartite, organization2022}. One such initiative is the Global Antimicrobial Resistance and Use Surveillance System (GLASS) \cite{GLASS}, which aims to allow standardisation of data collection, analysis, interpretation, and sharing, and encourages comprehensive surveillance on AMR through the collection of epidemiological, clinical, and population-level data. At present, multiple databases collect information on specific aspects of AMR, and in combination, inform surveillance, policy and research on AMR, primarily at the national level. The use of data technologies has the potential for an integrative analysis and interpretation of this data to better inform AMR control measures globally. \hfill

\subsection{Overview of data and sequencing technologies}

\noindent In the last few decades, data technologies such as artificial intelligence and machine learning have advanced greatly and are more commonly used as part of biomedical and healthcare research. However, they still have not made a significant impact on AMR research or been integrated in routine clinical practice.  With the increasing availability of large-scale and detailed datasets and ongoing efforts to standardize and harmonize information on AMR, data technologies will become an essential part of any analysis.\hfill\break\break
\noindent The main categories of data technologies are artificial intelligence, machine learning, and mathematical and statistical modelling. We define them below.\hfill\break\break 
Artificial intelligence (AI) is a broad term describing computational systems that attempt to accomplish specific or general tasks using algorithms. When these algorithms include the ability to infer patterns directly from data, without these patterns being explicitly specified, such systems are typically described as machine learning (ML), a term that also subsumes deep learning. Machine learning may be used, for instance, to use large datasets to learn patterns that endow a pathogen with antimicrobial resistance, or a small molecule with antimicrobial properties. \hfill\break\break
On the other hand, mathematical and statistical modelling efforts typically rely more heavily on the knowledge of one or more human experts to narrow down the possible patterns to be learned. Such models may be used, for instance, to predict the spread of an infection in the population or estimate the resulting morbidity, mortality and economic costs. However, both statistical modelling and ML follow closely related paradigms to extract meaningful patterns from datasets \cite{MLandSM}. \hfill\break\break
In parallel with data technologies, recent decades have seen an explosive growth in sequencing technologies, which has resulted in a rapid increase in the amount of sequencing data available. The use of next-generation sequencing (NGS) techniques, including whole-genome sequencing (WGS), is becoming more prevalent in AMR work, especially in surveillance and diagnostics. These techniques work by generating a large number of genome fragments, called reads, from each bacterial sample, which are then analysed to provide a detailed information about the underlying genome. Data technologies are then needed to turn these analyses into fast and accurate interpretation of the NGS data, such as the species and the resistance-determining variants present in the sample, the evolutionary relatedness of the organisms, and their potential sources \cite{mcdermott2021predicting}. \hfill\break\break
\noindent In this review we aim to synthesise available information on current usage of various data technologies in AMR research and discuss potential opportunities for further integration of these technologies in One Health research approach to combat the global AMR crisis.

\subsection{Previous work and our contribution}

\noindent Several other review papers have examined the role of AI and related data technologies in combating AMR \cite{AIDrugDiscovery, AIReviewAMR, Anahtar}. Most of these typically focus on a single area such as surveillance via genotype-phenotype prediction \cite{PredictionBestPractices}, diagnosis \cite{AIDiag}, and antibiotic discovery \cite{AIDrugDiscovery}, or explore AI applications in clinical decision support \cite{AIDecisionID} and infection management \cite{AIinID}. Our key contribution is to cover all aspects of AMR as well as to provide specific recommendations for action that can address the challenges currently being faced by data technologies in supporting AMR control measures. \hfill

\begin{figure}[ht]
    \centering
    \includegraphics[width=0.85\textwidth]{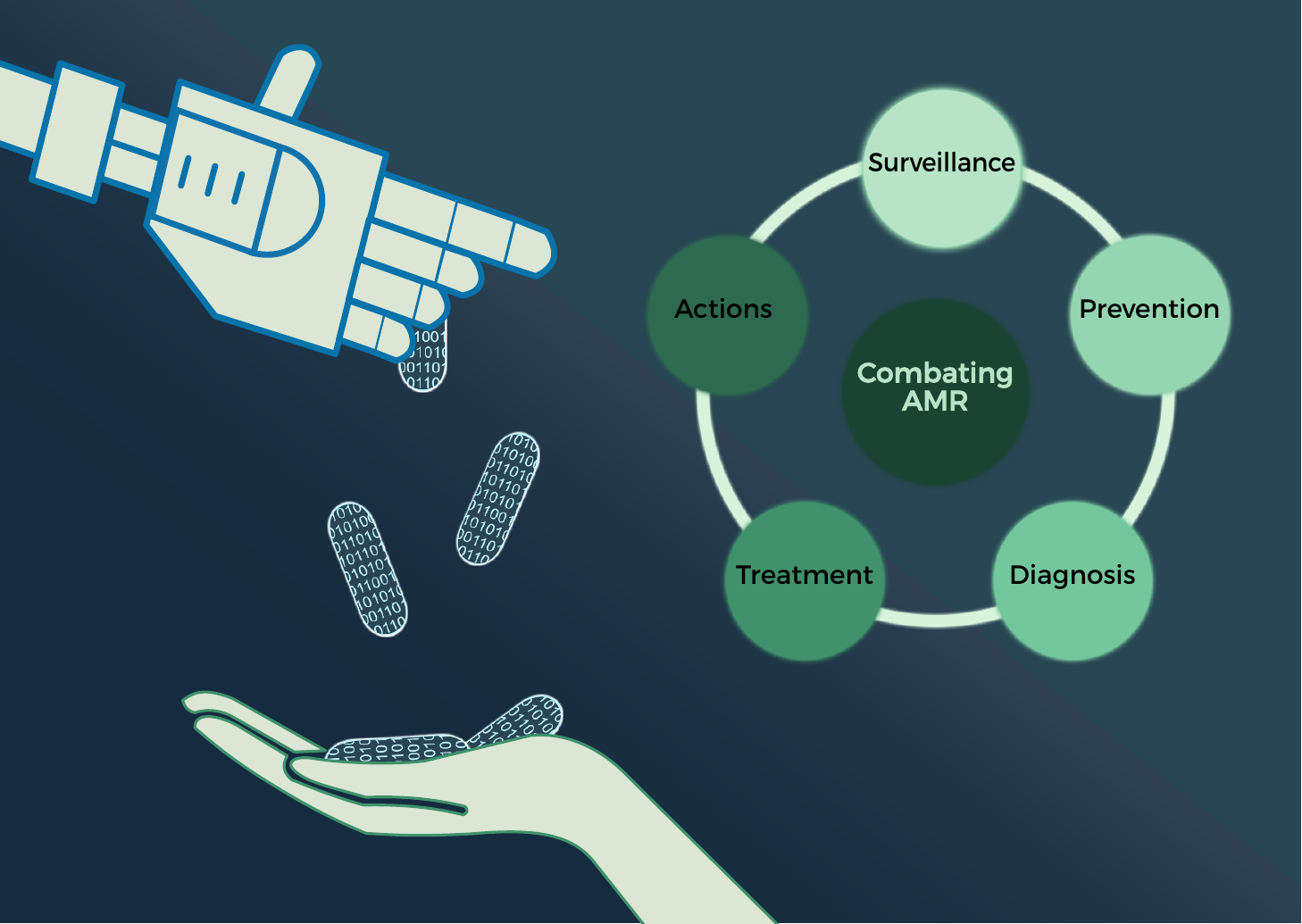}
    \caption{Conceptual overview of the paper's structure}
    \label{F:Overview}
\end{figure}

\noindent In this report, we summarise the current use of data technologies and discuss the potential for further integrated application of data technologies to combat AMR in each of the key areas of AMR research, as illustrated in Figure \ref{F:Overview}:

\begin{itemize}
 \item \textbf{Surveillance}: tracking AMR trends, observing unusual or new AMR patterns and alerting in case of concerning events such as outbreaks and emergence of specific AMR bacterial clones. 
 \item \textbf{Prevention}: interventions to prevent AMR infections, and stop the emergence and the spread of AMR.
 \item \textbf{Diagnosis}: detecting infections, the organisms causing them, and determining their AMR profiles (informative about suitable treatments).
 \item \textbf{Treatment}: optimizing treatment of infections to ensure a successful outcome, investigation of new treatment options, improving adherence.
\end{itemize}

\noindent We conclude it with a list of recommended actions to ensure the adoption and success of data technologies for combating AMR.

\section{Surveillance} \label{sec:surv}

\noindent Surveillance of AMR entails the systematic collection, aggregation, analysis, and timely dissemination of data on resistance patterns and antimicrobial prescribing and consumption. These types of surveillance systems are instrumental in developing evidence-based policies to control the spread of AMR and are essential to assess the impact of interventions to: (i) track antibiotic resistance and usage patterns over time and space; (ii) define and track genetic AMR variants and AMR bacteria; (iii) understand the relationship between antimicrobial use and the development of resistance; (iv) detect potential AMR outbreaks; and (v) attribute AMR to its source. 

\subsection{Tracking antimicrobial prescribing and resistance trends}

\noindent A cornerstone of the effort to reduce AMR is the surveillance of local and national antibiotic consumption and resistance rates in humans, animals, food, and the environment. Such surveillance data is used to examine the relationship between the two trends, track the impact of interventions on antibiotic usage and resistance rates, and inform decision-making at the clinical and policy level. \hfill\break\break
\noindent There are several national antibiotic prescribing and resistance surveillance programs; a fairly comprehensive inventory was put together by Diallo \textit{et al} \cite{SurvReview}, so we do not replicate it here. These surveillance programs employ independent strategies to collect data on AMR and antibiotic usage (AMU) rates, and rarely do so across the entire One Health spectrum. Furthermore, the reports sometimes become publicly available one or several years after data collection. The major strengths of such platforms and reports are that the data on antibiotic consumption and resistance is collected in a systematic manner, allowing direct comparison of data within the platform.\hfill\break\break 
However, differences exist across platforms in terms of the scale of data collected, the standards followed to determine resistance status (CLSI \cite{CLSI}, EUCAST \cite{EUCASTWEB}, BSAC \cite{BSACWEB}), the source of the sample (blood, sputum, stool), the type of infection (bloodstream, upper respiratory, urinary tract infection), the antibiotics included in the antimicrobial susceptibility testing profiles, and the population within the scope of their data collection (hospitals, community, animals). To overcome some of the challenges posed by inconsistencies across different antibiotics and AMR platforms and monitor AMR at a global level, the WHO proposed a collaborative effort to standardize AMR surveillance, GLASS \cite{GLASS}, that provides core protocols to ensure that the collected data will be valid and comparable across countries.\hfill\break\break
\noindent Integrated surveillance characterized by systematic collection of data across human, animal, food and environmental contexts allows for the development of effective AMR controls by recognizing the interconnected nature of AMR, varying evolutionary pressures and multiple exposure routes; see Figure \ref{F:Spheres}. This has been recognized in the WHO's One Health approach to AMR \cite{perez2015role,lammie2016antimicrobial}. When available, AMU and AMR data collected across the One Health spectrum can be used to assess the public health impact of antimicrobials that are used in humans and in the agricultural sectors, and to understand the genetic mechanisms used by different organisms to develop resistance. Consistently collected data allows for both country-specific and cross-country analysis.\hfill\break\break
Once interventions are put in place to control AMR, there may be a delay until the impact of these measures can be quantified, but high-quality longitudinal multi-country surveillance data will be key in estimating the effects of interventions. However, such successful, long-term, well-managed surveillance systems require a long period of investment and institutional commitment at the local, regional, national and global level that must withstand changes in politics and leadership.\hfill 

\subsection{Defining and tracking AMR bacteria and genetic AMR variants} \label{SS:AMR_gen}

\noindent AMR is typically determined based on antimicrobial susceptibility testing (AST) whereby minimum inhibitory concentrations (MICs) are interpreted using clinical guidelines \cite{CLSI, EUCASTWEB}, producing susceptible or resistant results. However, WGS techniques can also provide AST results, often referred to as WGS-AST, for instance by relying on one of the publicly available databases documenting known resistance mechanisms. A selection of such databases appears in Table \ref{T:Databases}.\hfill\break
\begin{table}[ht]
    \centering
    \begin{tabular}{|c|c|c|c|c|}
    \hline
    Name                & Host          &   Curation    & Assembled genomes & Ref.              \\ \hline
    CARD/RGI            & McMaster      &   \cmark      & \xmark            & \cite{CARD}       \\
    AMRFinderPlus       & NCBI          &   \cmark      & \cmark            & \cite{AMRFinder}  \\ 
    ARG-ANNOT           & UMR CNRS      &   \cmark      & \xmark            & \cite{ARGANNOT}  \\
    ResFinder           & DTU           &   \cmark      & \xmark            & \cite{ResFinder}  \\
    PATRIC              & NIAID         &   \cmark      & \cmark            & \cite{PATRIC}     \\
    ARGMiner            & Virginia Tech &  Crowdsourced & \xmark            & \cite{ARGMiner}   \\ \hline
    \end{tabular}
    \caption{A selection of resistance determinant databases.}
    \label{T:Databases}
\end{table}

\noindent While AST is the most common approach to quantify AMR levels via the MIC, it does not explain the routes or mechanisms of AMR acquisition, which WGS-AST can often provide information about. However, our limited knowledge of the genetic mechanisms of resistance and of their interaction within the same organism is an obstacle, and can be overcome to a degree by data technologies such as ML. In some bacterial species such as \textit{M. tuberculosis} and \textit{S. aureus}, WGS-AST is now as accurate as laboratory testing \cite{GenomicVsLab}, while other species (e.g. \textit{P. aeruginosa}) remain challenging \cite{PredChallenge}.\hfill\break\break
\noindent Substantial effort has been invested in developing ML algorithms that improve WGS-AST using the presence or absence of specific genes, single-nucleotide polymorphisms (SNPs), or genomic fragments known as $K$-mers. Other ML-based approaches include analysing conserved genes \cite{GenoOnly1}, matching isolates to their nearest genetic neighbor \cite{KNN}, or analysing genomic data in combination with gene expression data \cite{ExpressionAMR}. 
By following a training process described in more detail in the Diagnostics section, these algorithms can produce high-quality resistance predictions within a familiar collection of genomes, often meeting diagnostic standards \cite{Ferreira2020}, but may struggle to generalize to new isolates \cite{HicksGrad}, especially isolates coming from under-represented geographical settings or AMR mechanisms \cite{PredictionBestPractices}.\hfill\break\break
\noindent By incorporating WGS data, and measuring selective pressures such as AMU, surveillance programs can elucidate the routes of resistance evolution and transmission, allowing for a better design and evaluation of interventions. Understanding the drivers of AMR transmission, the role of international travel \cite{EuroTravelLaos}, and the degree of spillover of resistance between humans, animals, food and the environment \cite{Spillover, Ludden, duarte2021metagenomics} will assist in prioritising global efforts to reduce the threat of AMR to public health.

\subsection{AMR trend prediction and intervention models}

\noindent Having high-quality detailed longitudinal data on AMU as well as genetic characteristics of AMR pathogens across different regions would allow us to analyse and predict arising AMR trends to better understand associations between AMU and AMR pathogen strains and provide insight into where intervention measures could be taken. More studies are needed to demonstrate the added utility of these advanced data technologies and estimate the impact of proposed intervention measures. To date, the majority of mathematical models investigating AMR are concentrated on human infections \cite{niewiadomska2019population}. However, there is a need to develop integrative models that capture the interactions between humans, animals, and the environment, as well as the effects of vaccination and other AMR interventions, as the insights from such models could lead to more informed policy decisions \cite{knight2019mathematical, OneHealthModel}.\hfill

\section{Prevention}

\noindent Preventing the emergence and spread of AMR can be achieved by reducing the number of infections that occur and changing the pressures for resistance to emerge and spread among infected humans and animals. Data science and advancement of AI and ML techniques have allowed analysis of large-scale datasets to identify potential patterns and trends in the data, such as the emergence of carbapenem resistance \cite{CarbapenemResistance} and the spread of cefiderocol resistance \cite{Cefiderocol} in various pathogens. 

\subsection{Infection prevention and control (IPC) measures}

\noindent IPC is a complex, multifactorial endeavour, which makes it challenging to evaluate key interventions. Routine collection and real-time analysis of data across individual institutions is vital if currently available antibiotics are to be safeguarded. In clinical settings, personal hygiene, especially hand hygiene is considered the most important IPC measure \cite{Allegranzi}, but compliance with hand hygiene guidance can be low (around 40\%) \cite{Pittet, Erasmus} and has been identified as a key area in need of improvement to stop person-to-person transmission of AMR \cite{zingg2019technology}. Technological advancements to monitor hand hygiene technique and compliance and prompt hand hygiene behaviours have been found effective in increasing compliance \cite{Salman, Dyson}. IPC measures in healthcare settings also benefit from the rapid identification and characterisation of an infectious agent and its AMR profile, further discussed in the Diagnostics section.\hfill \break \break
\noindent Data technologies such as AI and ML are also being applied in food safety. By being able to more quickly detect outbreaks of food-borne diseases, and identify their source, public health officials can conduct recalls, alert consumers, and reduce the number of infections. WGS and ML also support food officials in identifying pathogen reservoirs, determining high-risk foods, and targeting interventions to prevent future infectious outbreaks \cite{GenomeTrakr}.

\subsection{Effective design and delivery of vaccines to combat AMR}

\noindent Vaccination is now considered a key part of the AMR strategy in many national action plans \cite{WHO_VACIM}. Vaccines help by preventing bacterial infections, thereby lessening antibiotic consumption, averting AMR infections, and reducing AMR levels in bacterial populations \cite{VaccinesAMR}. The same is true in the animal sector, where vaccines are widely use to prevent inflections and can also be used to reduce the prevalence of certain zoonotic pathogens, such as \textit{Salmonella} \cite{ZoonoticSalmonella}. While investment in antibiotic development is decreasing, investment and innovation in vaccine technologies are on the rise \cite{Vaccines, Azimi}.\hfill\break\break 
Designing an effective vaccination campaign can be a challenge in poorly mapped locations \cite{utazi2019mapping}. This design process can benefit from data technologies, such as satellite images that help map a country's population density and determine where vaccines need to be delivered. Thus, a partnership between Meta and the Red Cross used ML on billions of images to identify the best places to administer vaccines in Malawi in 2019 \cite{MetaRC2019}. Tracking the general population's uptake and acceptance of the vaccines (e.g. via medical records) can provide information on how messaging needs to change to appeal to the population.

\subsection{Identifying populations at highest risk for AMR infections}

\noindent Different pathogenic bacterial strains may pose different risks of causing severe infections in patients, depending on patient risk factors, antibiotic treatment regime and the virulence of the pathogen. Two main approaches are currently being explored for better stratifying patients by their individual risk profiles: analysis of bacterial phenotypic and/or genetic characteristics to predict outcome of infections, and analysis of patient’s clinical data, solely or in combination with bacterial characteristics. \hfill\break\break
Studies have reported bacterial genetic markers associated with pathogenicity and virulence for major families of pathogens \cite{Recker, Wheeler2018-pn, Van_Puyvelde2019-jh}. However, phenotype and genotype data from the pathogen, as well as clinical information, are needed for these analyses. These can cause challenges not only based on which data gets routinely collected, but also the volume of data needed to make accurate predictions to identify most vulnerable and at high-risk patients. ML algorithms have also been applied to patients’ electronic healthcare records to predict the individualised likelihood of patients developing sepsis \cite{wang2021machine}, outcomes form bloodstream infections \cite{zoabi2021predicting}, healthcare-associated infections \cite{barchitta2021machine} and \textit{C. difficile} infections \cite{panchavati2022comparative}, as well as detect ongoing outbreaks and save costs \cite{PittsburghStudy}.\hfill\break\break
\noindent Preventing additional AMR from emerging and spreading is key to ensuring other interventions continue to be successful after they are delivered. We have described some strategies for achieving this; other key factors are timely, accurate diagnosis and treatment of AMR infections, which we turn to next.\hfill

\section{Diagnosis} \label{sec:diag}

\noindent The misuse of broad-spectrum antibiotics, which can kill multiple types of bacteria, is one of the primary causes of AMR \cite{BroadSpectrum}. It is critical to use the right antibiotic at the right time to keep antibiotics that can treat drug-resistant infections viable. However, front-line practitioners must start antibiotic therapy right away if a patient presents a risk of serious infection such as bacteraemia (bloodstream infection), which may lead to sepsis \cite{SepsisOverview}. Selecting the initial antibiotic is crucial, and is done with a limited capacity to foresee AMR, resulting in the overuse of broad-spectrum antibiotics \cite{BroadSpectrum}. \hfill\break\break
\noindent A key reason for the use of broad-spectrum antibiotics is the lack of rapid point-of-care (PoC) diagnostic tests \cite[Rapid Diagnostics Chapter]{ONeill}. Traditional diagnostic approaches rely on a central laboratory to carry out microbiological, molecular biological, or serological techniques \cite{Alternatives}. As an alternative, AI-based algorithms can often analyze the patient's clinical, demographic and electronic health record (EHR) data on the spot \cite{AIDiag}. They may identify a bacterial infection by its characteristic patterns, look for validated human biomarkers, or use a patient's EHR and clinical symptoms to assess their risk of a bacterial infection. This process can support the clinician's diagnosis and treatment decisions and facilitate the prescription of more targeted antibiotics \cite{Prescription}.\hfill

\begin{figure}
    \centering
    \includegraphics[width=0.78\textwidth]{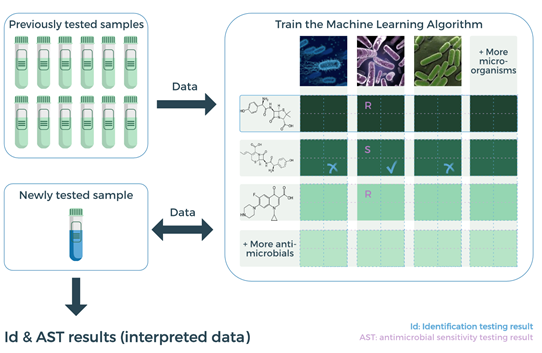}
    \caption{Schematic diagram of a typical ML approach for pathogen identification and susceptibility determination.}
    \label{F:Diag}
\end{figure}

\subsection{Current status of data technologies in AMR diagnosis}
\noindent ML-based clinical decision support systems can leverage a patient's electronic health record to assess their individual risk of a resistant infection and recommend the narrowest-spectrum antibiotic likely to successfully treat the infection \cite{Anahtar}. A number of trials have tested simple rule-based clinical decision support systems that do not employ ML \cite{voermans2019cost}. Some ML-based systems have shown clear potential to reduce over-prescribing of broad-spectrum antibiotics in prospective trials \cite{TSystems, chowdhury2020identifying}, while others have yet to progress to such trials \cite{moran2020towards, kanjilal2020decision}.\hfill\break\break
\noindent In 2020, 222 AI-based medical devices were registered in the USA \cite{Muehlematter}, and 64 of them were FDA-approved \cite{Benjamens}; in the EU, where approval of medical devices is more decentralised \cite{EMARules}, 240 AI-based approved medical devices were approved \cite{Muehlematter}. Most of these devices are used for diagnostic purposes, but only 4 of them target microbiology -  2 approved both in the USA and Europe (one for sepsis and one for automated plate reading) and 2 only in the USA (one for sepsis and one for syphilis) \cite{Muehlematter}. Most AI algorithms used for diagnostics are based on pattern recognition \cite{Benjamens, Muehlematter}. These algorithms can process large volumes of data in a short time, detect subtle patterns in this data, and facilitate detection of disease, including infectious disease \cite{AIDiag}.\hfill\break\break
Broadly speaking, AI approaches for the diagnosis of an infection (Dx), the identification of an infecting organism (Id), and the determination of its AST profile can be categorized into two groups: those that involve genome sequencing, reviewed in \cite{EUCASTReview, NGSChallenges} and discussed briefly in the Surveillance section, and those that do not, reviewed in \cite{OrganismID, Kasas2021, Kaprou2021}. They are summarised in Table \ref{T:Diags}. \hfill

\subsection{Challenges for data technologies in AMR diagnosis}
\noindent As illustrated in Figure \ref{F:Diag}, the development of the diagnostic tool requires two components. First, a sufficiently large collection of existing examples of input data (data available at point-of-use, typically, genomic data, a biological or biochemical screen, and/or EHRs) and outcome data (Dx, Id or AST) needs to be collected. This database serves as the training data for the AI algorithm to learn diagnosis-, organism-, or antibiotic-specific patterns.\hfill\break\break
\noindent Second, the algorithm needs to be trained on this database to extract patterns sufficient to distinguish between outcomes of interest: bacterial infections vs non-bacterial (such as viral or fungal) infections or other diseases (Dx), possible infectious bacteria (Id), or susceptible and the resistant versions of an organism (AST). This training process involves tuning the model to optimise its accuracy, and validating it on previously unseen data.\hfill\break\break
\noindent Among the key challenges faced by all of these approaches is the requirement to assemble a sufficiently large, representative, and diverse database to train the AI algorithm and to identify the optimal training methodology. While genomics-based approaches can benefit from a large number of publicly available sequences \cite{COBS}, these are not always accompanied by high-quality AST profiles \cite{Ferreira2020}, while other technologies tend to require specific experimental data that may be kept proprietary for competitive reasons \cite{ASTDatabases}.\hfill

\begin{longtable}[c]{|M{0.389\linewidth}|M{0.311\linewidth}|M{0.03\linewidth}|M{0.03\linewidth}|M{0.045\linewidth}|M{0.21\linewidth}|}
    \hline Data source or technique & Example tool [Ref.] & Dx & Id & AST & Availability \\\hline
    \endfirsthead
    \multicolumn{6}{|c|}{Genomic Approaches}\\\hline
    Metagenomics + mapping    & Patho-Seq, Ripseq \cite{Pathogenomix} & \xmark    & \cmark    & \xmark    & In development
    \\
    Metagenomics + mapping    & MIST \cite{MIST} & \xmark    & \cmark    & \xmark    & Open source
    \\ 
    Metagenomics + mapping  & ThermoFisher Ion16S \cite{Ion16S} & \xmark    & \cmark    & \xmark    & On the market 
    \\
    Nanopore + custom pipeline       & Grumaz \textit{et al} \cite{Fraunhofer}  & \cmark    & \cmark    & \xmark    & In development
    \\
    Nanopore + alignment   &  Schmidt \textit{et al} \cite{NanoporeUrine} & \xmark    & \cmark    & \cmark    & In development
    \\
    Nanopore + assembly     & Taxt \textit{et al} \cite{Nanopore} &  \xmark    & \cmark    & \cmark    & In development
    \\
    Padlock probe detection & Mezger \textit{et al} \cite{Padlock} & \xmark    & \cmark    & \cmark    & In development
    \\
    PCR + high-resolution melt  & Athamanolap \textit{et al} \cite{PCRMelt} & \xmark    & \cmark    & \cmark    & In development
    \\
    Random DNA probes       & UMD \cite{UMD} & \xmark    & \cmark    & \xmark        & In development
    \\
    Transcriptomic signatures   & Reyes \textit{et al} \cite{Transcriptomics} & \cmark    & \xmark    & \xmark    & In development
    \\
    WGS + specialised probes & IDbyDNA Explify \cite{IDbyDNA} & \cmark    & \cmark    & \cmark    & Research use
    \\
    WGS + $K$-mer similarity & DayZero epiXact \cite{DayZero} & \xmark  & \cmark    &   \cmark       & On the market
    \\
    WGS + custom pipeline & Ares Genetics ARESiss \cite{Galata2019} & \xmark    & \cmark    & \cmark    & On the market
    \\
    \hline
    \multicolumn{6}{|c|}{Non-genomic Approaches}\\\hline
    App + disk diffusion    & Antibiogo \cite{Pascucci2020} & \xmark    & \xmark    & \cmark    & Open source 
    \\
    Automated microscopy & Accelerate Pheno \cite{Accelerate} & \cmark  & \cmark    &   \cmark  & On the market
    \\
    Clinical data + EHR  & TREAT \cite{TSystems} & \cmark  & \cmark    &   \xmark       & On the market
    \\
    Clinical data + EHR  & Navoy Sepsis \cite{SepsisDiag} & \cmark  & \xmark    &   \xmark       & In development
    \\
    Clinical data + EHR  & Lewin-Epstein \textit{et al} \cite{IsraeliML}  & \xmark    & \xmark    & \cmark    & In development
    \\
    Electrochemical profiles & MicroplateDx \cite{Hannah2020} & \cmark    & \xmark    & \cmark    & In development
    \\
    Flow cytometry           & FASTinov \cite{Costa2017} & \xmark  & \xmark    &   \cmark       & On the market
    \\
    Flow cytometry           & FAST \cite{FlowCytoAST} & \xmark  & \xmark    &   \cmark       & Research use
    \\
    Gas chromatography + volatile organic compounds & Imspex BreathSpec \cite{VOC} & \cmark  & \cmark    &   \xmark       & In development
    \\
    Growth kinematics  & QuickMIC \cite{Wistrand2020} & \xmark    & \xmark          &   \cmark       & In development
    \\
    Growth plate assessment & Clever Culture APAS \cite{APAS} & \xmark    & \xmark    &   \cmark      & On the market
    \\
    Impedance cytometry & iFAST \cite{iFAST} & \xmark    & \xmark    & \cmark        & In development
    \\
    Leukocyte profiling     & Coulter DxH800 \cite{Coulter} & \cmark    & \xmark        & \xmark    & On the market
    \\
    MALDI-TOF mass spec  & VITEK MS \cite{Vitek} & \cmark  & \cmark    &   \cmark       & On the market
    \\
    MALDI-TOF mass spec  & Bruker Biotyper \cite{Bruker} & \cmark  & \cmark    &   \cmark       & Research use
    \\
    MALDI-TOF mass spec  & DRIAMS \cite{DRIAMS} & \xmark  & \xmark    &   \cmark       & In development
    \\
    Metabolism + fluorescence   & SNDA-AST \cite{Avesar2017} & \xmark    & \xmark    & \cmark    & In development
    \\
    Microcalorimetry     & Symcel calScreener \cite{Symcel} & \xmark  & \xmark    &   \cmark       & On the market
    \\
    Microtiter plate + LED array    & AST Reader \cite{Microtiter} & \xmark    & \xmark    & \cmark    & In development
    \\
    Nanomotion           & Resistell \cite{Kasas2021}  & \xmark    & \xmark    & \cmark    & In development
    \\
    Raman spectroscopy   & Nostics \cite{Nostics} & \cmark  & \cmark    &   \xmark     & In development
    \\
    Single-cell phenotyping  & PhAST \cite{PHaST} & \cmark  & \cmark    &   \cmark       & In development
    \\
    \hline
    \caption{Some diagnostic tools leveraging AI or ML.}
    \label{T:Diags}
\end{longtable}

\section{Treatment}

\noindent While prevention, surveillance and diagnosis have a critical role to play, the development of new treatments, including both traditional antimicrobial drugs and combinations as well as alternative approaches such as bacteriophages or other personalised therapies, continues to be the mainstay of infectious disease treatment. Furthermore, the use of data technologies to ensure treatment adherence may help address a contributor to AMR: the failure to complete a full course of antibiotic treatment, although some studies question its role in AMR \cite{FailureToComplete}.

\subsection{New antimicrobials}

\noindent To reduce the burden of AMR, immediate action needs to be taken to develop new antibiotics \cite{Wright}. So far, most antibiotics have been discovered by exploring naturally occurring compounds, for example by screening soil-dwelling microbes for antibacterial chemicals \cite{NaturalProducts}. Unfortunately, in the last four decades, few antibiotics and no new classes of antibiotics have been developed \cite{Pipeline}. A 2017 estimate found that it would cost 1.5 billion US dollars to develop a new antibiotic, but that it would only generate 46 million US dollars in revenue, explaining the market dropouts \cite{Enne}. Changes to the incentive structures have been suggested to encourage the development of antibiotics, with the UK now trying a subscription-based pricing model \cite{Incentives}.\hfill\break\break
AI offers a different approach to the problem by trying to make the screening of candidate antibiotic molecules more efficient and less costly. Its uses in drug discovery more broadly are reviewed in \cite{EndToEndAI, AIForDrugDiscovery, AIForDrugDiscoverySequel} and in antibiotic discovery specifically, in \cite{AIDrugDiscovery, SustainableAbx}. Algorithms can be helpful in several aspects of the process: they may help identify compounds with a desired mechanism of action or target (efficacy) \cite{Efficacy}, anticipate potential toxicity problems (safety) \cite{AdmeTox}, and optimise their uptake by the human body (pharmacokinetics) \cite{PKPD}.\hfill\break\break
\noindent An important step in using AI to discover new candidate antibiotics is to represent potential molecules in a machine-readable way, for instance by using tools from cheminformatics \cite{EndToEndAI}. Once molecules are represented in this way, an algorithm can be trained to recognize if a molecule has antibiotic properties by comparing it to other antibiotics (positive examples) and non-antibiotics (negative examples). The vector representation can be created manually by providing a list of quantitative properties for each molecule (feature engineering) or learned automatically from a string or graph representation of the molecule, as done in deep learning \cite{MolFeatures}.\hfill\break\break
\noindent A recent success of ML was the identification of a compound, halicin, that could inhibit the growth of a number of challenging bacteria, including \textit{C. difficile} and pan-resistant \textit{A. baumannii} \cite{DLCollins}. An initial library of compounds tested for efficacy against \textit{E. coli} was represented using a deep learning method and molecular properties. The trained model was applied to over 6000 previously tested drug candidates \cite{DRH} to computationally identify those with high antibacterial efficacy and low toxicity. 
A different way of leveraging data technologies, specifically mathematical modelling, is demonstrated in prioritising the chemical derivatives of macrolide antibiotics \cite{Derivatives}. Their common chemical structure was explored to derive a 
model of both the molecule in isolation as well as in a state of bound to the bacterial ribosome. 
This work led to the discovery of a compound with an easier synthetic route, a higher solubility and binding affinity to the ribosome, and over 50 times more potency against several priority resistant bacterial pathogens, including \textit{A. baumannii}, \textit{P. aeruginosa}, and \textit{E. coli}.

\subsection{Combination therapies}

\noindent Combination therapies, which involve the joint administration of two or more antibiotics \cite{Doern}, are another potential solution to reduce the development of AMR in the context of a limited drug discovery pipeline \cite{Fischbach, Chait}. The identification of suitable combination partners and prediction of combination effectiveness of existing drugs is a well-studied ML problem in other disease areas such as cancer \cite{Anchang, Malyutina}. Predicting the effectiveness of antibiotic combinations currently relies on a number of parameters, including the intrinsic mutation rate of bacterial pathogens, the pharmacokinetic and pharmacodynamic parameters of the agents under study \cite{Niu, Jacobs}, and pharmacogenomic data suggesting risk factors for toxicity or drug interactions \cite{Vardakas, Apter}. The Antimicrobial Combination Network \cite{Jorge} is a database evaluating many of these parameters.\hfill\break\break
\noindent A specific mathematical model developed to assess antibiotic combinations, DiaMOND \cite{Cokol}, first quantifies all pairwise combinations of antibiotics, then assesses a small number of informative high-order combinations to characterise the landscape of possible combinations against \textit{M. tuberculosis}. Another method \cite{DosageCombo} uses an optimisation algorithm to predict the optimal dosage of two antibiotics against carbapenem-resistant \textit{A. baumannii}, validated in a hollow fiber model of infection \cite{HollowFiber}. Machine learning is the basis of several other approaches; CoSynE \cite{Cosyne} is a workflow to learn a model of antibiotic interactions from chemical structure descriptors, while CARAMeL \cite{Caramel} learns these interactions by leveraging metabolic network models of \textit{E. coli} and \textit{M. tuberculosis}; both methods correctly predict many novel interactions.

\subsection{Antimicrobial peptides and other alternative approaches}

\noindent Antimicrobial peptides (AMPs) are short chains of amino acids; they occur naturally as part of the immune system \cite{AMPReview}. AMPs provide an alternative to antibiotics since they exhibit potent antibacterial and anti-biofilm activity by acting on multiple targets in the plasma membrane and intracellular components of pathogenic bacteria \cite{Zhang}. Computational tools, including ML and deep learning, play a significant role in AMP identification (in both prokaryotic and eukaryotic organisms) and design (\textit{de novo}, i.e. from scratch) \cite{WangDeep}, although only a handful of cationic AMPs have so far been approved for clinical use by successful clinical trials \cite{AMPTrials}.\hfill\break\break
\noindent The identification problem for AMPs includes efforts to identify biosynthetic gene clusters (BGCs) in archaea and bacteria as well as naturally occurring peptides in the mammalian proteome, an area of research often referred to as genome mining \cite{GenomeMining}. Several databases cataloguing existing AMPs support these efforts, of which the Antibiotics and Secondary Metabolites Analysis Shell (antiSMASH) \cite{antiSMASH} and the Database of Antimicrobial Activity and Structure of Peptides (DBAASP) \cite{Pirtskhalava} are prominent examples. Promising approaches include both methods using simple models or scoring functions \cite{NPLinker} as well as deep learning models \cite{AMPlify}.\hfill\break\break
\noindent Concurrently with the development of many alternatives to antibiotics and AMPs, data technologies continue to play an important role in the prioritisation of promising candidates. In particular, ML approaches have shown promising results in predicting each of the following: the success or failure for a faecal microbiota transplantation (FMT) based on both donor and recipient microbiome profiles \cite{kazemian2020trans}, the effectiveness of a bacteriophage cocktail based on a joint analysis of bacterial and bacteriophage omics information \cite{Lood2022}, the effect of nanoparticles against AMR bacteria alone \cite{Mirzaei} and in combination with existing antibiotics \cite{DieguezSantana2021}, and the effect of probiotics based on an individual's characteristics \cite{Montassier}.

\subsection{Treatment adherence and monitoring}

\noindent Lack of adherence to antibiotic therapy leads to poor clinical outcomes and increased risk of treatment failure and AMR \cite{Adherence}. Smart medication adherence products (MAPs), defined as devices that provide real-time medication intake data that can be stored and analyzed automatically, may help alleviate this risk \cite{Faisal}. For instance, the MediSafe app has been shown to improve adherence to antibiotic treatment in adults with irritable bowel syndrome \cite{MediSafe}. \hfill\break\break 
ML can additionally predict the likelihood that a patient would follow through on their prescribed course of antibiotics, thus helping to focus resources on the most at-risk patients; a recent study achieved an accuracy above 80\% in identifying those patients who had less than 50\% chance of purchasing a prescribed drug \cite{IsraeliAdherence}. Lastly, automated alert systems based on an algorithm can notify clinicians about at-risk patients in a timely manner by a direct analysis of their EHR data; a digital sepsis alert was shown to increase timely administration of antibiotics, and reduce the risks of long hospital stay and death, in a London hospital \cite{DigitalSepsis}.\hfill

\section{Recommendations for using data technologies against AMR} \label{sec:rec}

\noindent The approaches to reducing the burden of AMR outlined in the previous sections not only rely on innovations in data technologies, but also effective implementation of such technologies and straight-forward integration within the current already existing systems. Here we outline key recommendations to ensure these strategies are implemented and are implemented effectively. 

\subsection{Data collection}

\noindent Our first two recommendations focus on data collection practices; specifically, we recommend collecting AMR data with relevant metadata in standardised formats and making it machine-readable to facilitate analysis.\hfill\break\break
\noindent \textbf{Recommendation 1}: Collect and report AMR data together with rich metadata in a well-defined format.\hfill\break\break
\noindent Metadata, defined as ``data that provides information about other data'', is a key component of data collection efforts. For AMR surveillance data this can include the date, the location, and the source of reporting; for AMU data, the prescribing authority, the indication, and any relevant diagnostic tests carried out; for AMR diagnostics, the technology used to process the isolate (which also includes the sequencing platform in the case of WGS-based approaches), relevant laboratory conditions, and the agreement between replicates. The presence of relevant metadata, especially if it is in a clear and well-defined format, can facilitate the smooth integration of disparate data sources and help account for biases in the results.Of course such data collection still needs to ensure patient confidentiality (if data includes patient attributable data). \hfill\break\break
\noindent \textbf{Recommendation 2}: Ensure access to machine-readable data for relevant stakeholders (clinical teams, public health professionals and similar).\hfill\break\break
\noindent Some of the data relevant to AMR is found in formats that are poorly readable (e.g. scanned handwritten laboratory reports), inconsistently structured (e.g. electronic health records), or containing proprietary elements not easily processed by free software tools (e.g. surveillance reports using the PDF or some of the Microsoft Office suite formats). All of these present additional access barriers and data integration challenges, as much of the data processing time is spent on cleaning the data and harmonizing the formats \cite{EHR}, which might still result in those cleaned databases not being easily accessible. Despite advances in technologies such as optical character recognition (OCR) and natural language processing (NLP), a systematic analysis of AMR data is well-served by the use of machine-readable, open formats during the data collection and collation stage.\hfill

\subsection{Data privacy considerations}

\noindent Our next two recommendations focus on the tension between controlling AMR and respecting data privacy, especially in data collected from patients or owned by commercial entities (including animal agriculture and food production companies) with an interest in keeping it proprietary.\hfill\break\break
\noindent \textbf{Recommendation 3}: When possible, obtain consent for patient sample use in research.\hfill\break\break
\noindent Whenever the use of personal data in AMR control is considered, this must be done under the auspices of data protection and privacy laws. Irrespective of the national and international variations within privacy laws, such as the General Data Protection Regulation (GDPR) in Europe and the Health Insurance Portability and Accountability Act (HIPAA) in the USA \cite{Nurmi, SciPol}, the primary purpose of these laws is to protect the privacy of the individual, including but not limited to their identity, personal characteristics and health status. As patient samples (e.g. blood or urine) may provide invaluable information on causative pathogen, AMR mechanisms, transmission, and treatment efficacy, but may still contain patient cells and DNA, it is desirable to obtain consent for their use for research purposes.\hfill\break\break
\noindent \textbf{Recommendation 4}: Use federated learning to analyse AMR data from multiple sources.\hfill\break\break
\noindent While many ML algorithms require large training datasets, privacy considerations may make data sharing a challenge. The WHO's GLASS initiative \cite{GLASS}, as well as other programs (e.g. ATLAS \cite{ATLAS}), are taking steps towards creating representative collections of pathogen isolates (both resistant and sensitive) for surveillance and diagnosis, but currently available collections remain small \cite{PredictionBestPractices}. Federated learning is a paradigm in which multiple collaborators train a ML model, each on their own data, and send their model updates to a central trusted server to be aggregated into a consensus model \cite{FederatedML}; this may help improve model accuracy without compromising privacy. Similarly, the discovery of new antimicrobials via ML requires exploring compound databases, many of which are proprietary to biopharmaceutical companies; federated learning may provide opportunities to share these databases without revealing sensitive proprietary information \cite{FederatedDrugs}. An alternative model, which has met with some success in other therapeutic areas, is an open platform for sharing chemical and biological screen information; a noteworthy effort along those lines is SPARK (Shared Platform for Antibiotic Research and Knowledge) by the Pew Trust \cite{SPARK}.
\hfill

\subsection{Data technology development}

\noindent Our next recommendations are concerned with the reliability of the models or conclusions arising from the data and provide two best practice suggestions derived from related fields.\hfill\break\break
\noindent \textbf{Recommendation 5}: When possible, use gold standard AMR data with a faithful representation of the underlying uncertainty to train predictive models and algorithms.\hfill\break\break
\noindent An algorithm can never produce an output of higher quality than that of its input data. The use of a gold standard counteracts the danger of low-quality input data by increasing the chance that the input data is free of errors (such as misdiagnosed cases); for clinical outcomes such as treatment success, this could be obtained from a consensus of experts \cite{Consensus}. The gold standard may have limited precision; for instance, the determination of an MIC by the common method of serial dilutions is accurate to within one dilution \cite{DilutionMIC}, so it can help for the representation of the results to reflect that uncertainty; for instance, representing the outcome as a range of $0.5$ to $1$ rather than a value of $1$ can lead to more stable MIC prediction algorithms. 
\hfill\break\break
\noindent \textbf{Recommendation 6}: Validate ML algorithms on an independent or benchmark AMR dataset.\hfill\break\break
\noindent Even with a gold standard, concerns often arise with the presence of unwanted bias in the data; for instance, certain geographical regions, strain types, or AMR sources may be over- or under-represented. Although every reasonable effort must be put into quality control of the data, the best practice for establishing the validity of a predictive algorithm is to validate it on a dataset independent from the one used to develop it. If an independent dataset is not available, the best practice is to estimate the performance of the algorithm on previously unseen data by \textit{cross-validation} \cite{CV}. This is a technique whereby the data is split into groups randomly, the algorithm learns from all the groups except one, and predicts on the remaining group. For AMR surveillance and diagnostics, some authors recommend using groups of closely related strains, not random groups, for more realistic estimates \cite{PySeer}. \hfill\break\break
\noindent Benchmark datasets have been used in areas of ML such as image classification \cite{BenchmarkCV}, and natural language processing \cite{BenchmarkNLP} to ensure a common standard against which all the algorithms' performance is measured. To the authors' knowledge, with the recent exception of Raphenya \textit{et al} \cite{Raphenya} there is no currently accepted benchmark dataset for any of the tasks related to AMR, such as the prediction of AMR incidence or AMR phenotypes, the identification of AMR mechanisms, or the estimation of a candidate molecule's antimicrobial properties. Such benchmark datasets would require effort and resources to curate, and inevitably suffer from limitations \cite{BenchmarkLimits}, but would serve as an invaluable resource for algorithm developers by allowing for consistent performance comparisons.\hfill

\subsection{Data technology dissemination}

\noindent Our final two recommendations focus on the use of the data technologies and ensuring their continued usefulness and relevance for AMR control efforts.\hfill\break\break
\noindent \textbf{Recommendation 7}: Design AMR data systems with semantic interoperability in mind.\hfill\break\break
\noindent Semantic interoperability is the ability of computer systems to exchange data with unambiguous, shared meaning \cite{SemanticInter}. Studies have shown that approaches rooted in semantic interoperability, such as the ones adopted by the WHO Collaborating Centre for Surveillance of Antimicrobial Resistance (WHONET) \cite{WHONET} and the Antimicrobial Resistance Trend Monitoring System (ARTEMIS) \cite{ARTEMIS}, increase the efficiency and reliability of online AMR monitoring systems. The harmonisation of systems reporting AMR predictions from genomic data is currently being undertaken by the Public Health Alliance for Genomic Epidemiology (PHA4GE) \cite{PHA4GE}. This recommendation complements our first recommendation on rich AMR metadata in a well-defined format. \hfill\break\break
\noindent \textbf{Recommendation 8}: Contribute to creating, curating and updating AMR databases.\hfill\break\break
\noindent Effective policies for the use of data technology against AMR should encourage academics, healthcare providers and industry participants to share their data. Curated public repositories of available data that are accessible to any interested party upon reasonable request are needed, not only for benchmarking purposes (Action 6), but also to lower the barrier to entry for new contributors. Currently available genome databases include those hosted by the National Center for Biotechnology Information (NCBI), the European Molecular Biology Laboratory (EMBL), and the DNA Data Bank of Japan (DDBJ) \cite{SRA}, as well as pathogen-specific databases such as Pathosystems Resource Integration Center (PATRIC) \cite{PATRIC} and the National Microbial Pathogen Data Resource (NMPDR) \cite{NMPDR}. A unified interface for accessing these disparate genomic databases, like the GISAID database used for SARS-CoV-2 genomes \cite{GISAID}, can be a helpful tool.\hfill\break\break 
\noindent Similarly, isolate banks can be an invaluable resource for AMR diagnostics and drug developers. Currently available pathogen isolate banks include the FDA-CDC Antimicrobial Resistance Isolate Bank \cite{FDACDC}, the Culture Collection of the University of Gothenburg, Sweden \cite{CCUG}, which collaborates with EUCAST \cite{EUCASTWEB}, and the National Collection of Type Cultures \cite{NCTC} at the UK Health Security Agency. However, some of these may be biased towards historic rather than modern isolates, omit representatives of the most recently evolved resistance mechanisms, and require substantial resources for curation and maintenance. With the right incentives for contributors, a globally representative strain bank could be a preparedness solution for an AMR pandemic.\hfill

\section*{Conclusion}

\noindent In this report we have provided an overview of current data technologies and their use to help reduce the burden of AMR globally. However, as some of these technologies are still mainly used for research purposes, there are steps needed to integrate these technologies in other areas, as explained in the Recommendations section. To progress towards a better integration and implementation of data technologies in different sectors to help track and prevent AMR, and allow a transparent, confident sharing of data or results between different platforms and institutions, investments are needed to ensure that core components of the foundation data systems, including surveillance systems, commercial and government data systems, population data systems and data from academic research, are properly set up in the respective institutions, ensuring that quality assurance and regular verifications are taking place. This together with the specified data collections will enable the judicious use of data technologies to control AMR.\hfill\break\break
\noindent At the global level, technical standards for data technology and systems development are also imperative to ensuring that data technologies are effectively embedded and used within these systems. To this end, a global focus on data stewardship, exchange of data and greater technical guidance focused on the implementation of data technologies in low- and middle- as well as high-income settings is important. In addition, due to the global nature of the AMR control challenges, open sharing of the data fuelling and underlying the data technologies to the extent possible under the constraints of relevant commercial and intellectual property interests and privacy considerations should be a key policy focus.\hfill\break\break
Every country will have a different context and organisational model for implementing data technologies for AMR, and a ``One Health'' approach to foundation data systems will help identify how data technologies are best prioritised and invested in as assets within the system. A roadmap and business plan may assist policymakers to identify how data technologies can be implemented as part of their broader data systems \cite{GHRU}. As part of this systems approach, regular updating of the data and processing of feedback on how the data is being used will ensure that stakeholders are aware of the value of the use of data technologies to control AMR.\hfill

\section*{Acknowledgements}

\noindent The authors would like to acknowledge Dr Frances Davies for assistance with the clinical accuracy of the Future of AMR Control section and Vivien van Dongen for designing two visualisations for this report. \hfill\break\break
LC acknowledges funding from the MRC Centre for Global Infectious Disease Analysis (reference MR/R015600/1), jointly funded by the UK Medical Research Council (MRC) and the UK Foreign, Commonwealth \& Development Office (FCDO), under the MRC/FCDO Concordat agreement and is also part of the EDCTP2 programme supported by the European Union. EJ is an Imperial College Research Fellow jointly supported by the Rosetrees Trust and the Stoneygate Trust (M683) and is affiliated with the National Institute for Health Research Health Protection Research Unit (NIHR HPRU) in Healthcare Associated Infections and Antimicrobial Resistance at Imperial College London in partnership with the UK Health Security Agency (previously PHE), in collaboration with Imperial Healthcare Partners, University of Cambridge and University of Warwick.\hfill\break\break
The views expressed are those of the authors and not necessarily those of the NIHR, Public Health England, the Department of Health and Social Care, or other organisations the authors are affiliated with.\hfill

\section*{Author contributions}

\noindent Conceptualisation of overall report and supervision of the working groups: LC, EJ, HG, MvD. Investigation, original draft writing and editing, by section: Surveillance  - EJ, NW, ANg, JR; Prevention - EJ, NW, KA, BA, LM; Diagnostics - LC, DB, MG, YK, ANg, LU; Treatment - LC, WA, KF, CM, AN, KR; Recommended actions - VL, SN, AS, DW, MvD. Additional visualisation: KF. Writing of the overall original draft: LC, EJ, NW. Critical review of the final draft: SB, GG, TR, RS. All co-authors have read and approved the final version of the manuscript. \hfill

\section*{Declaration of Interests}

\noindent LC is a consultant with Pfizer and the Foundation for Innovative New Diagnostics. MG and HG are a co-owners of a consulting business in the life sciences. RS is a co-founder and Chief Scientific Officer of PhAST Diagnostics, an infectious diseases diagnostic company. VL has held an investigator-initiated grant through Pfizer. MvD is the director of AMR Insights BV, a globally active, network-based information provider on antimicrobial resistance. All other co-authors have nothing to declare.

\section*{List of abbreviations used}
\begin{table}[h]
    \centering
    \begin{tabular}{ll|ll}
    AI      & artificial intelligence   &   MGE     & mobile genetic element \\
    AMP     & antimicrobial peptide     &   MIC     & minimum inhibitory concentration \\
    AMR     & antimicrobial resistance  &   ML      & machine learning \\
    AMU     & antimicrobial use         &   NGS     & next-generation sequencing \\
    AST     & antimicrobial susceptibility testing  &       SNP     & single-nucleotide polymorphism\\
    IPC     & infection prevention and control      &       WGS     & whole-genome sequencing \\
    \end{tabular}
    \label{T:Abbr}
\end{table}

\clearpage

\bibliographystyle{elsarticle-num}

\end{document}